\documentclass[utf8]{frontiersSCNS} 

\usepackage{url,hyperref,lineno,microtype,subcaption}
\usepackage{float}

\def\keyFont{\fontsize{8}{11}\helveticabold }
\def\firstAuthorLast{Pan Zihan {et~al.}} 
\def\Authors{Pan Zihan\,$^{1}$, Chua Yansong\,$^{2*}$, Wu Jibin\,$^{1}$, Zhang Malu\,$^{1}$, Li Haizhou\,$^{1}$, and Eliathamby Ambikairajah\,$^{3}$}

\def\Address{
$^{1}$ Department of Electrical and Computer Engineering, National University of Singapore, Singapore, 117583 \\
$^{2}$ Institute for Infocomm Research Agency for Science, Technology and Research, Singapore, 138632 \\
$^{3}$  School of Electrical Engineering and Telecommunications, University of New South Wales, Australia}

\def\corrAuthor{Corresponding Author}
\def\corrEmail{James4424@gmail.com}

\begin{document}
\onecolumn
\firstpage{1}

\title[An efficient auditory neural encoding]{An efficient and perceptually motivated auditory neural encoding and decoding algorithm for spiking neural networks}

\author[\firstAuthorLast ]{\Authors}

\address{} 

\correspondance{} 

\extraAuth{}

\maketitle

\begin{abstract}

Auditory front-end is an integral part of a spiking neural network (SNN) when performing auditory cognitive tasks. It encodes the temporal dynamic stimulus, such as speech and audio, into an efficient, effective and reconstructable spike pattern to facilitate the subsequent processing. However, most of the auditory front-ends in current studies have not made use of recent findings in psychoacoustics and physiology concerning human listening. In this paper, we propose a neural encoding and decoding scheme that is optimized for speech processing. The neural encoding scheme, that we call Biologically plausible Auditory Encoding (BAE), emulates the functions of the perceptual components of the human auditory system, that include the cochlear filter bank, the inner hair cells, auditory masking effects from psychoacoustic models, and the spike neural encoding by the auditory nerve. We evaluate the perceptual quality of the BAE scheme using PESQ; the performance of the BAE based on speech recognition experiments. Finally, we also built and published two spike-version of speech datasets: the Spike-TIDIGITS and the Spike-TIMIT, for researchers to use and benchmarking of future SNN research.

\tiny
 \keyFont{ \section{Keywords:} Spiking neural network, neural encoding, auditory perception, spike database} 
\end{abstract}

\section{Introduction}
\label{sec: introduce}

The temporal or rate based Spiking Neural Networks (SNN), supported by stronger biological evidence than the conventional artificial neural networks (ANN), represents a promising research direction. Neurons in a SNN communicate using spiking trains that are temporal signals in nature, therefore, making SNN a natural choice for dealing with dynamic signals such as audio, speech, and music.

In the domain of rate-coding, we studied the computational efficiency of SNN \citep{pan2019neural}. Recently, further evidence has supported the theory of temporal coding with spike times. To learn a temporal spike pattern, a number of learning rules have been proposed, which include the single-spike Tempotron  \citep{gutig2006tempotron}, conductance-based Tempotron \citep{gutig2009time}, the multi-spike learning rule ReSuMe \citep{ponulak2010supervised} \citep{taherkhani2015dl}, the multi-layer spike learning rule SpikeProp \citep{bohte2002error}, and the Multi-spike Tempotron  \citep{gutig2016spiking}, etc. The more recent studies are aggregate-label learning \citep{gutig2016spiking}, and a novel probability-based multi-layer SNN learning rule (SLAYER) \citep{shrestha2018slayer}.


In our research, we are constantly addressing the question: what are the advantages of SNN over ANN? From the viewpoint of neural encoding, we expect to encode a dynamic stimulus into spike patterns, which was shown to be possible \citep{maass1997networks} \citep{ghosh2009spiking}. Deep ANNs have benefited from the datasets created in recent years. In the field of image classification, there is ImageNet \citep{deng2012imagenet} \citep{ILSVRC15}; in the field of image detection, there is COCO dataset \citep{veit2016coco}; while in the field of Automated Speech Recognition (ASR), there is TIMIT for phonemically and lexically transcribed speech of American English speakers \citep{garofolo1993timit}. With the advent of these datasets, better and faster deep ANNs inevitably follow \citep{simonyan2014very}\citep{hochreiter1997long}\citep{redmon2016you}. The publicly available datasets become the common platform for technology benchmarking. In the study of neuromorphic computing, there are some datasets such as N-MNIST \citep{orchard2015converting}, DVS Gestures \citep{amir2017low} and N-TIDIGITS \citep{anumula2018feature}. They are designed for SNN benchmarking. However, these datasets are relatively small compared to the deep learning datasets.

One may argue that the benchmarking datasets for deep learning may not be suitable for SNN studies. Let us consider image classification as an example. Humans process static images in a similar way as they would process live visual inputs. We note that live visual inputs contain much richer information than 2-D images. When we map \citep{rueckauer2017conversion} or quantize \citep{zhou2016dorefa} static images into spike trains, and compare the performance of an ANN on static images, and a SNN on spike trains, we observe an accuracy drop. One should however not hastily conclude that SNNs are inherently poor in image classification as a consequence of event-based activations in SNNs. Rather, the question seems to be: how can one better encode images into spikes that are useful for SNNs, and how can one better use these spikes in an image classification task? For some of the recent image-based neuromorphic datasets, Laxmi et al \citep{iyer2018neuromorphic} has argued that no additional information is encoded in the time domain that is useful for pattern classification. This prompts us to look into the development of event-based datasets that inherently contain spatio-temporal information. On the other hand, a dataset has to be complex enough such that it simulates a real-world problem. There are some datasets that support the learning of temporal patterns \citep{wu2018biologically}\citep{zhang2018highly} \citep{zhang2017empd} \citep{MPDAL},  whereby each pattern contains only a single label, such as a sound event or an isolated word. Such datasets are much simpler than those in deep learning studies \citep{graves2006connectionist} \cite{graves2012sequence}, whereby a temporal pattern involves a sequence of labels, such as continuous speech. For SNN study to progress from isolated word recognition towards continuous speech recognition, a continuous speech database is required. In this paper, we would describe how we convert the TIMIT dataset to its event-based equivalent: Spike-TIMIT. 

A typical pattern classification task consists of three stages: encoding, feature representation, and classification. The boundaries between each stage are getting less clear in an end-to-end classification neural network. Even then, a good encoding scheme can significantly ease the workload of the subsequent stages in a classification task, for instance, the Mel-Frequency Cepstral Coefficients (MFCC) \citep{mermelstein1976distance} is still very much in use for automatic speech recognition (ASR). Hence the design of a spiking dataset should consider how the encoding scheme could help reduce the workload of the SNN in a classification task. This cannot be misconstrued as giving the SNN an unfair advantage so long as all SNNs are measured using the same benchmark. The human cochlea performs frequency filtering \citep{tobias2012foundations} while human vision performs orientation discrimination \citep{appelle1972perception}. These all involve encoding schemes to help us better understand our environment. In our earlier work \citep{pan2019neural}, on a simple dataset TIDIGITS \citep{leonard1993tidigits} that contains only single spoken digits, we used a population threshold coding scheme to encode the dataset into events, which we refer to as Spike-TIDIGITS. Using such an encoding scheme, we go on to show that the dataset becomes linearly separable, i.e., the input can be classified based on spike counts alone. This demonstrates that when information is encoded in both the temporal (spike timing) and spatial (which neuron to spike) domain, the encoding scheme is able to project the inputs to a higher dimension, that takes some of the workload off the subsequent feature extraction and classification stages. In the case of Spike-TIDIGITS, the spikes encoded can be directly counted and then classified using a Support Vector Machine (SVM). Using this neural encoding scheme, We further enhance it and then apply it to the TIMIT dataset in this work.

The motivation of this paper is two-fold. Firstly, we believe that we need well-designed spike-encoded datasets that represent the state-of-the-art encoding methodology. With these datasets, one can focus the research on SNN feature representation and classification tasks. Secondly, the datasets should present a challenge in pattern classification, that become the reference benchmark in future SNN studies.

As speech is the most common way of human communication, we are looking into the neural encoding of speech signals in this work. The first question is how best  possible to convert speech signals into spikes. As it is, there have been many related works in speech and audio encoding, each of which is optimized for a specific objective, for example, minimum signal reconstruction error  \citep{xiao2016spiking} \citep{dennis2013temporal} \citep{loiselle2005exploration}. However, none of them is optimized for  neuromorphic implementation, that considers the psycho-acoustics, computational efficiency, and effectiveness for pattern classification. In the SNN applications for speech recognition \citep{xiao2016spiking} \citep{darabkh2018efficient}, Mel-Frequency Cepstral Coefficients (MFCC)  \citep{mermelstein1976distance} are commonly used as the spectral representation in speech recognition. Others have tried to use the biologically plausible cochlear filter bank, but they are either analog filters which are prone to changes in the external environment \citep{liu2010neuromorphic}, or yet to be studied in a spike-driven SNN system \citep{loiselle2005exploration}. 

Considering spectral representation, an important step in neural encoding is to then convert the spectral energy in a perceptual frequency band into a spike train.  The most common way is to treat the two-dimensional time-frequency spectrogram as a static image, then converting each 'pixel' value into a spike latency time within the framing window size \citep{wu2018biologically}, or into the phase of the sub-threshold membrane potential oscillation \citep{nadasdy2009information}. Such methods do not represent the spatio-temporal dynamics of the auditory signals in a way that can be directly learned in a SNN \citep{wu2018spiking}. Furthermore, these prior studies mostly encode all the frequency components in the frames, and all of these frames into spike trains, introducing a lot of redundancy and hence unnecessary computational load for the subsequent SNN processing, such as speech recognition. Finally, little research  has been studied on how to reconstruct a neural encoded speech signal back into its auditory signals for perceptual evaluation. Speech signal reconstruction is a critical task in speech information processing, such as speech synthesis, singing synthesis, and dialogue technology.

To address the need of neuromorphic computing for speech information processing,  we propose three criteria for a biologically plausible auditory encoding (BAE) front-end:

\begin{enumerate}[(1)]
    \item Biologically plausible spectral features.
    \item Sparse and energy-efficient spike neural coding scheme.
    \item Friendly for temporal learning algorithms on cognitive tasks.
\end{enumerate}

The fundamental research problem in neural encoding is how to encode the dynamic and continuous speech signals into discrete spike patterns. Spike rate code is thought to be less likely in an auditory system since much evidence suggest otherwise, such as the example of how bats rely highly on the precise spike timing of their auditory system to locate sound sources by detecting a time difference as short as $5\mu s$. Latency code and phase code are well supported by neuro-biological observations. However, on its own, they cannot provide an invariant representation of the patterns for a classification task. To facilitate the processing of a SNN in a cognitive task, neural temporal encoding should not only consider how to encode the stimulus into spikes, but also care about how to represent the invariant features. Just like the auditory and visual sensory representations in the human prefrontal cortex, such representations in the proposed BAE front-end are required in a SNN framework, that can then be implemented with a low cost neuromorphic solution, that can effectively reduce the processing workload in the subsequent SNN pipeline. A large number of observations in neuroscience support the observation that our auditory sensory neurons encode the input stimulus using threshold crossing events in a population of sensory neurons \citep{ehret1997auditory} \citep{hopfield2004encoding}. Inspired by these observations, a simple version of threshold coding has been proposed  \citep{gutig2009time}, in which a population of encoding neurons with a set of uniformly distributed thresholds encode the spectral energy of different frequency channels into spikes. Such a cross-and-fire mechanism is reminiscent of quantization from the point of view of information coding. In our proposed BAE encoding front-end, such a neural coding scheme is also being incorporated. Further investigation is presented in the experiment section. 

Besides effective neural coding representation, an efficient auditory front-end aims to encode acoustic signals into sparse spike patterns, while maintaining sufficient perceptual information. To achieve such a goal, our biological auditory system has provided us a solution best understood as masking effects\citep{harris1979forward}\citep{shinn2008object}. Masking is a complex and yet to be fully understood psychoacoustic phenomenon as some components of the acoustic events are not perceptible in both frequency and time domain \citep{ambikairajah1997auditory}. From the viewpoint of perceptual coding, these components are regarded as redundancies since they are inaudible. Implementing the masking effects, those inaudible components will be coded with larger quantization noise or not coded at all. Although the mechanism and function of masking is not yet fully understood, its effects have already been successfully exploited in auditory signal compression and coding \citep{ambikairajah2001wideband}, for efficient information storage,  communication, and retrieval. In this paper, we propose a novel idea to apply the auditory masking effects in both frequency and time domain, which we refer to as simultaneous masking and temporal masking, respectively, in our auditory neural encoding front-end so as to reduce the number of encoding spikes. This improves the the sparsity and efficiency of our encoding scheme. Given how we address the three optimization criteria of neural encoding, we refer to it as biologically plausible auditory encoding scheme or BAE. Such an auditory encoding front-end also provides an engineering platform to bridge the study of masking effects between psycho-acoustics and speech processing. 


Our main contributions in this paper are: 1) we emphasize the importance of spike acoustic datasets for SNN research. 2) we propose an integrated auditory neural encoding front-end to further research in SNN-based learning algorithms. With the proposed BAE encoding front-end, the speech datasets can be converted into an energy-efficient, information-compact, and well-representative spike patterns for subsequent SNN tasks.

The rest of this paper is organized as follows: in Section \ref{methods} we discuss the auditory masking effects, and how simultaneous masking in the frequency domain, and temporal masking in the time domain for neural encoding of acoustic stimulus is being implemented; the BAE encoding scheme is applied in conjunction with masking to TIDIGITS and TIMIT datasets. In Section \ref{experiment and results}, we describe the details of the resulting spike datasets and evaluate them in comparison with their original datasets in a recognition task. In Section \ref{discussion}, we discuss our findings and conclude in Section \ref{conclusion}.

\section{Materials and methods}
\label{methods}

\subsection{Auditory masking effects}
\label{auditory masking}

Most of our speech processing front-ends employ a fixed feature extraction mechanism, such as MFCC, to encode the input signals, whereas the human auditory sensory system ignores some while strongly emphasizes others, commonly referred to as attention mechanism in psycho-acoustics. The auditory masking effects closely emulate this phenomenon \citep{shinn2008object}. 

Auditory masking is a known perceptual property of the human auditory system that occurs whenever the presence of a strong audio signal makes its neighborhood of weaker signals inaudible, both in the frequency and time domain. One of the most notable application of auditory masking is the MPEG/audio international standard for audio signal compression \citep{fogg2007mpeg} \citep{ambikairajah2001wideband}. It compresses the audio data in large part by removing the acoustically irrelevant parts of the audio signal, or by encoding those parts with less number of bits, due to more tolerance to quantization noise \citep{ambikairajah1997auditory}. To achieve such a goal, this algorithm designs two different kinds of maskings from the psycho-acoustic model \citep{lagerstrom2001design}:
\begin{enumerate}
	\item In the frequency domain, two kinds of masking effects are used. Firstly, by allocating the quantization noise in the least sensitive regions of the spectrum, the perceptual distortion caused by quantization is minimized. Secondly, an absolute hearing threshold is exploited, below which the spectral components are entirely removed.
	\item In the time domain, the masking effect is applied such that the local peaks of the temporal signals in each frequency band will make their ensuing audio signals inaudible. 
\end{enumerate}

Motivated by the above signal compression theory, we propose an auditory masking approach to spike neural encoding, which greatly increases the coding efficiency of the spike patterns, by eliminating those perceptually insignificant spike events. The approach is conceptually consistent with the MPEG-1 layer III signal compression standard \citep{fogg2007mpeg}, with modifications according to the characteristics of spiking neurons. %

\subsubsection{Simultaneous masking}
\label{simultaneous masking}

The masking effect present in the frequency domain is referred to as simultaneous masking. According to the MPEG-1 standards, there are two sorts of masking strategies in the frequency domain: the absolute hearing threshold and the frequency maskers. The simultaneous masking effects are common in our daily life. For instance, the sensible sound levels of our auditory systems vary in different frequencies, therefore, we can be more sensitive to the sounds in our living environment. This is an evolutionary advantage for survival, in both human beings and animals. Besides the absolute hearing threshold, every acoustic event in the spectrum will also influence the perception of the neighboring frequency components, that is, different levels of tones could contribute to masking effects of other frequency tones. For instance, in a symphony show, the sounds from different musical instruments can be fully or partially masked by each other. As a result, we can enjoy the compositions of various frequency components with rich diversities. Such a psycho-acoustic phenomenon is called frequency masking.

\begin{figure}[H]
    \centering
    \includegraphics[width=0.7\columnwidth]{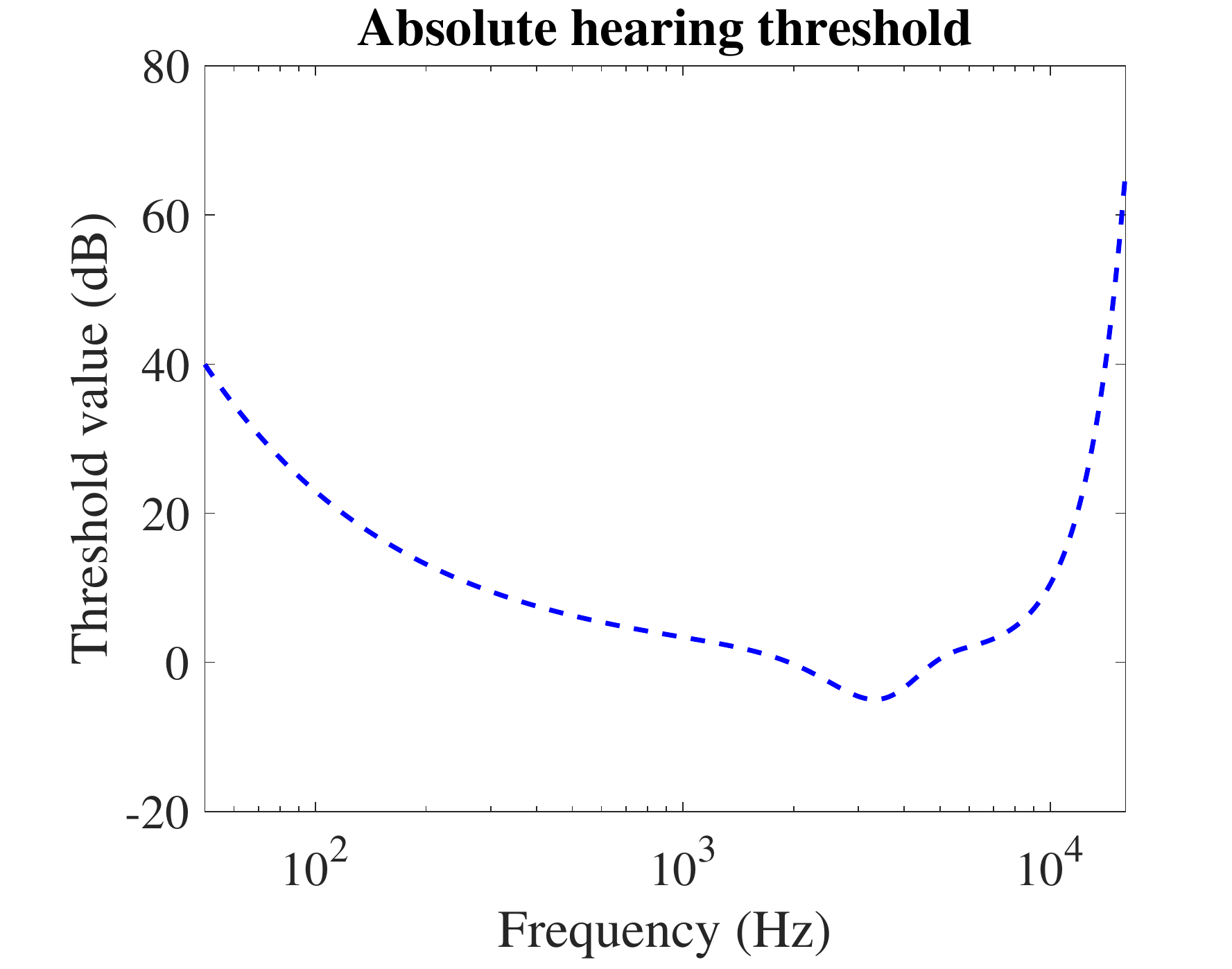}
    \caption{Absolute hearing threshold $T_a$ for the simultaneous masking. Our hearing is more sensitive to the acoustic stimulus around several thousand  Hz, that covers the majority of the sounds in our daily life. The sounds below the thresholds are completely inaudible.}
    \label{hearing threshold}
\end{figure}

Figure \ref{hearing threshold} illustrates the absolute hearing threshold, $T_a$, as a function of frequency in Hz. The function is derived from psycho-acoustic experiments, in which pure tones continuous in the frequency domain are presented to the test subjects and the minimal audible sound pressure levels (SPL) in dB are recorded. The commonly used function to approximate the threshold is \citep{ambikairajah1997auditory}:
\begin{equation}
	T_a(f)=3.64\times(\frac{f}{1000})^{-0.8}-6.5\times e^{-0.6(\frac{f}{1000}-3.3)^2}+0.001 \times (\frac{f}{1000})^4
\end{equation}

\begin{figure}[H]
    \centering
    \includegraphics[width=0.7\textwidth]{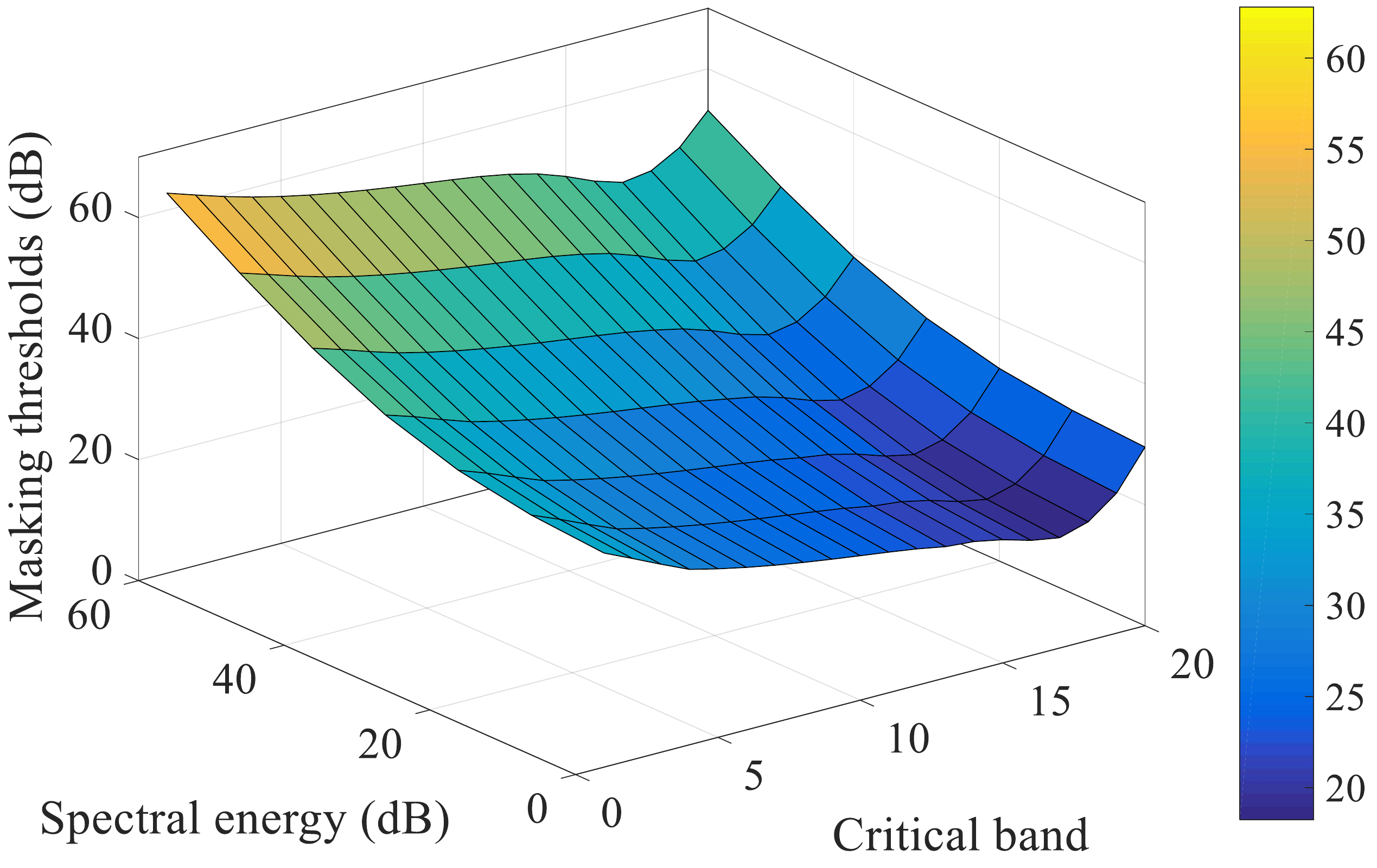}
    \caption{The frequency masking thresholds acting on a maskee (the acoustic events being masked), generated by the acoustic events from the neighboring critical bands, are shown as a surface in a 3-D plot. The acoustic events are referred to as the spectral power of the frames in a spectrogram. The spectral energy axis is the sound level of a maskee; the critical band axis is the frequency bins of the cochlear filter bank, as introduced in Section \ref{sec:introduce datasets}; the masking thresholds axis indicates the overall masking levels on the maskees of different sound levels from various critical bands. For example, an acoustic event of $20$dB level on the $10^{th}$ critical band is masked off by the masking threshold of nearly $23$dB, which is introduced by the other auditory components of its neighboring frequency channels.}
    \label{Fre maskers}
\end{figure}

For the frequency maskers, in the MPEG-1 standard, some sample pulses under masking thresholds might be partially masked, thus they are encoded by less number of bits. However, in the event-based scenario, spike patterns carry no amplitude information, similar to on-off binary values, which means that partial masking can hardly be realized. As such, we have modified the approach such that all components under the frequency maskers are fully masked (discarded). Further reconstruction and pattern recognition experiments are necessary to evaluate such an approach. Figure \ref{Fre maskers} shows the overall masking thresholds with both masking strategies in the frequency domain. This figure illustrates the simultaneous masking thresholds added to the acoustic events in a spectrogram. The sound signals with different spectral power in different cochlear filter channels will suffer from various masking thresholds.

\begin{figure}[H]
    \centering
    \includegraphics[width=0.7\textwidth]{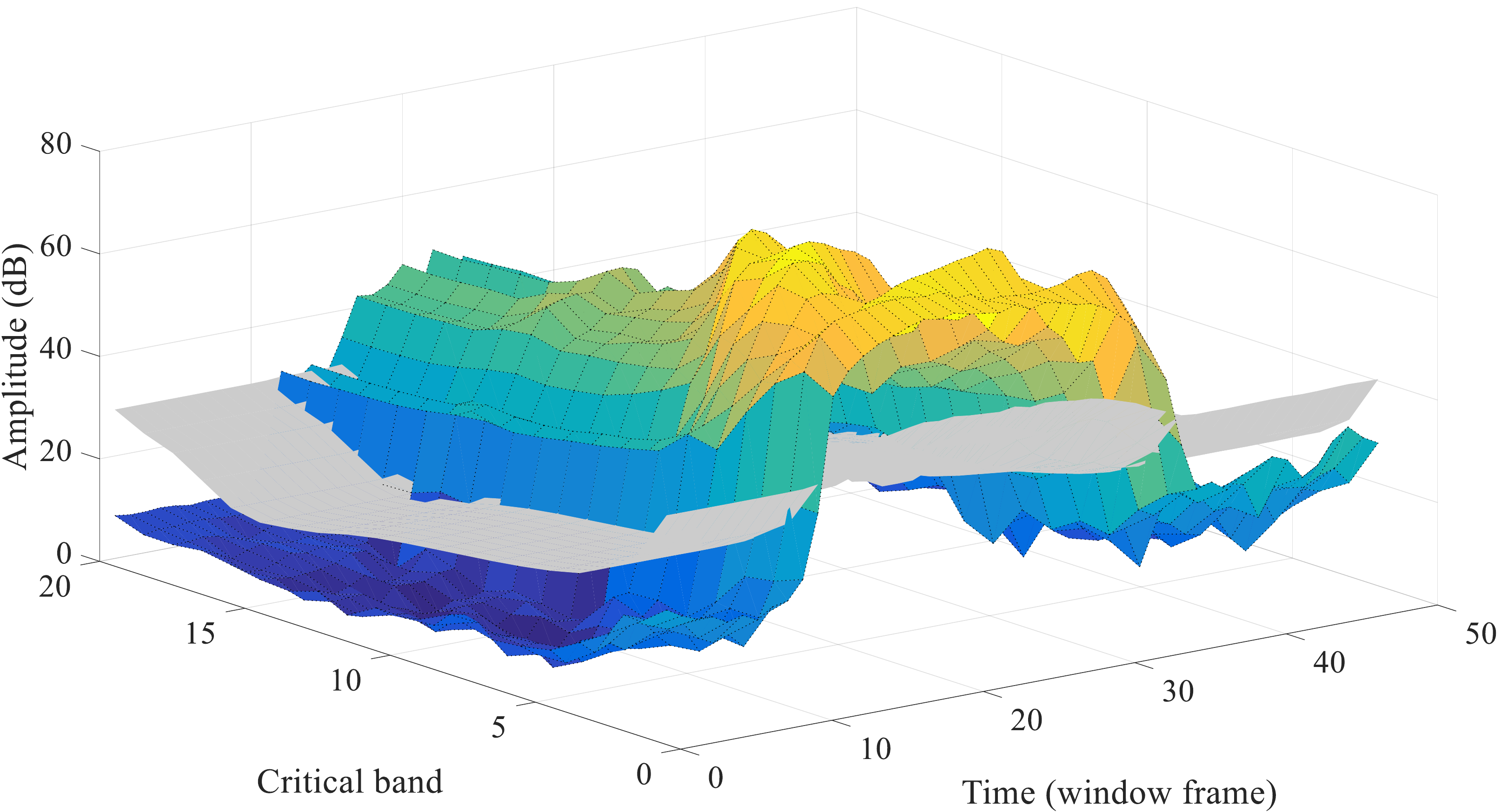}
    \caption{The overall simultaneous masking effects on a speech utterance of ``one'', in a 3D spectrogram. Combining the two kinds of masking effects in the frequency domain (refer to Figure \ref{hearing threshold} and Figure \ref{Fre maskers}), the grey surface shows the overall masking thresholds on a speech utterance (the colorful surface). All the spectral energy under the thresholds will be imperceptible.}
    \label{Fre mask 3D}
\end{figure}

Figure \ref{Fre mask 3D} provides a real-world example of the simultaneous masking. The spectrogram of a speech utterance of ``one'' from TIDIGITS dataset is demonstrated in a 3-D plot. The grey surface illustrates the simultaneous masking threshold acting on the spectrogram (colorful surface). By the masking strategy, the acoustic events with spectral energy lower than the threshold surface will be removed. Section \ref{sec: integrate encoder} will introduce how to convert the masked spectrogram into a sparse and well-represented spike pattern.

\subsubsection{Temporal masking}
\label{temporal masking}

\begin{figure}[H]
    \centering
    \includegraphics[width=0.6\textwidth]{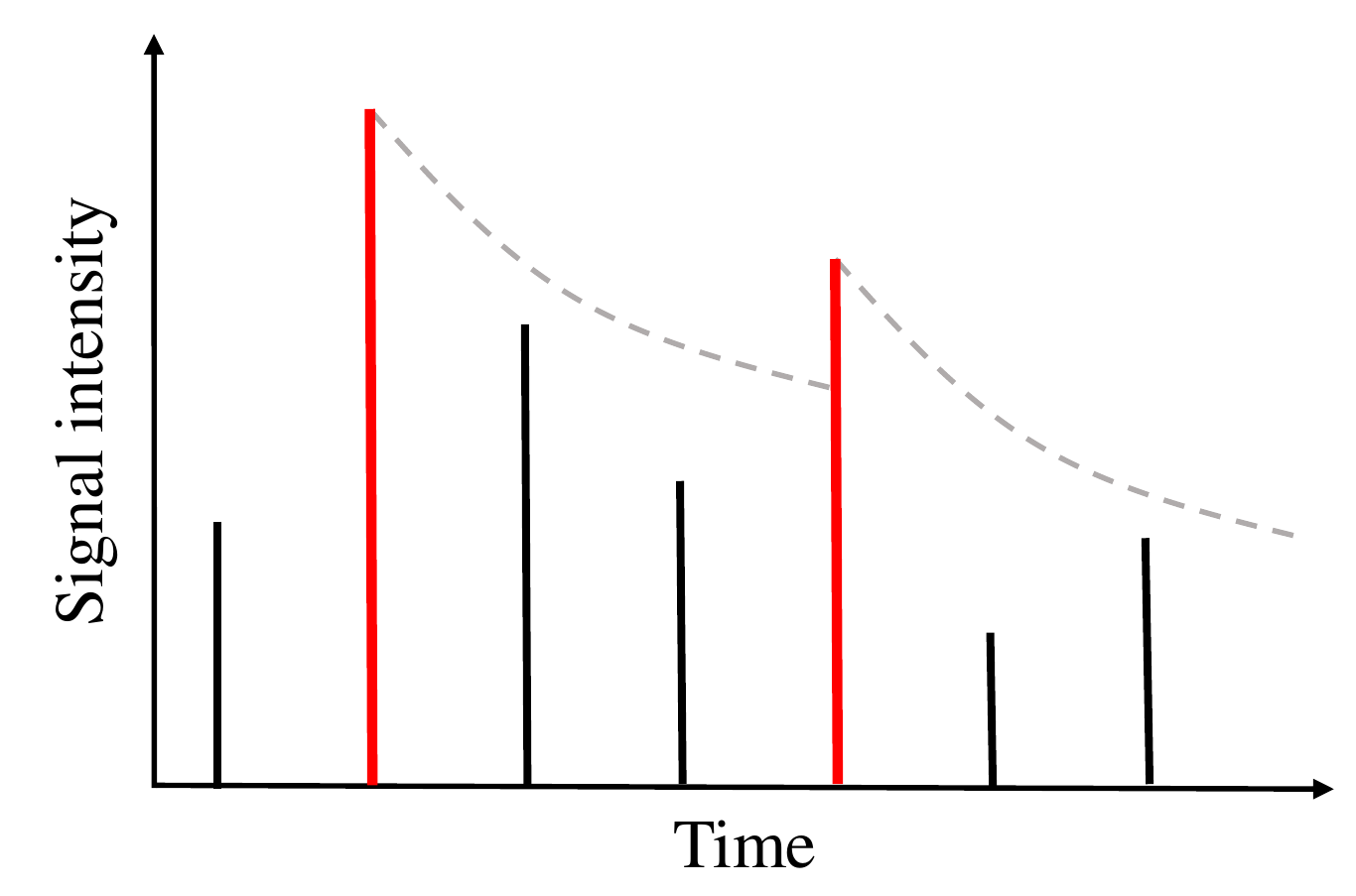}
    \caption{The illustration of temporal masking: each bar represents the acoustic event received by the auditory system. In this paper, acoustic events generally referred to framing spectral power, which are the elements to be parsed to an auditory neural encoding scheme. A local peak event (red bar) forms a masking shadow represented by an exponentially decaying curve. The subsequent events that are weaker than the leading local peak will not be audible until another local peak event exceeds the masker curve.}
    \label{Temp mask}
\end{figure}

Another auditory masking effect is temporal masking in the time domain. Conceptually similar to the frequency maskers, a louder sound will mask the perception of the other acoustic components in the time domain. As illustrated in Figure \ref{Temp mask}, the vertical bars represent the signal intensity of short-time frames, that is called acoustic events, along the time axis. A local peak (the first red bar) forms a masker that makes the following events inaudible until the next local peak (the second red bar) exceeds the masker curve. According to the psycho-acoustic studies, the temporal masker threshold is modeled as an exponentially decaying curve \citep{ambikairajah2001wideband}:

\begin{equation}
    y(n)=c^{n} \times p_1 
    \label{eq: temporal mask}
\end{equation}
where $y(n)$ denotes the masking threshold level on the $n^{th}$ following an acoustic event; $c$ is the exponential index and $p_1$ represents the sound level of the local peak as the beginning of the masker. The decaying parameter $c$ is tuned according to the hearing quality.


\subsubsection{Auditory masking effects in both domains}
\label{sec:masking in both domains}

By applying both the simultaneous masking and temporal masking illustrated above, we can remove those imperceptible acoustic events (frames) from the overall spectrogram. Since our goal is to apply the masking effects in the precise timing neural code, we propose the strategy as follows:

\begin{enumerate}
	\item The spike pattern $\mathbf{P}_{K \times N}(p_{ij})$ is generated from the raw spectrogram $\mathbf{S}_{K \times N}(s_{ij})$ without masking effects, by some temporal neural coding methods, which will be discussed in Section 2.2.2. Here the index $i,j$ refers to the time-frequency bin in the spectrogram, with $i$ referring to the frequency bin, and $j$ referring to the time frame index. The spike pattern $\mathbf{P}_{K \times N}$ is defined as a matrix that:
\begin{equation}
p_{ij}=
\begin{cases}
t_f, \quad \text{if a spike is emitted within the duration of the time-frequency bin $i,j$} \\
~0, \quad \text{otherwise} \\
\end{cases}
\end{equation}
where $t_f$ is the encoded precise spike timing. As such the spike pattern $\mathbf{P}_{K \times N}(p_{ij})$ is a sparse matrix that records the spike timing.
	
	\item According to the spectrogram  $\mathbf{S}_{K \times N}(s_{ij})$ and the auditory perceptual model, the simultaneous masking level matrix $\mathbf{M_\text{simultaneous}}(m^{\text{simultaneous}}_{ij})$ and the temporal masking level matrix $\mathbf{M_\text{temporal}}(m^{\text{temporal}}_{ij})$ are obtained. The overall masking level matrix $\mathbf{M}_{K \times N}(m_{ij})$ is defined as follows. It provides a 2-D masking threshold surface that has the same dimensions as the spectrogram.
	\begin{equation}
	    m_{ij}=\text{min} \left\{ m^{\text{simultaneous}}_{ij}, m^{\text{temporal}}_{ij} \right\}
	\end{equation}
	
	\item A masker map $\Phi_{K \times N}({\phi}_{ij})$ is generated, whose dimensions are the same as the spectrogram. The element of the matrix $\Phi_{K \times N}({\phi}_{ij})$ is defined as:
	
\begin{equation}
{\phi}_{ij}=
\begin{cases}
1, \quad \text{if $s_{ij} \geq m_{ij}$} \\
0, \quad \text{if $s_{ij} < m_{ij}$} \\
\end{cases}
\label{eq: masking map}
\end{equation}

where the time-frequency bin ${i,j}$ is masked with $\phi_{i,j} = 0$ when the frame energy $s_{ij}$ is less than the masking threshold $m_{ij}$, otherwise, $\phi_{i,j} = 1$.

	\item Apply the masker map matrix $\Phi_{K \times N}({\phi}_{ij})$ to the encoded pattern $\mathbf{P}_{K \times N}(p_{ij})$ to generate a masked spike pattern $\mathbf{P}^{\text{mask}}(p^{\text{mask}}_{ij})$ :
	\begin{equation}
		\mathbf{P}^{\text{mask}}=\mathbf{P}_{K \times N}\circ\Phi_{K \times N}
	\end{equation}
	where $\circ$ denotes the Hadamard product. By doing so, those perceptually insignificant spikes are eliminated, thus forming a more compact and sparse spike pattern.
\end{enumerate}

\begin{figure}[H]
    \centering
    \includegraphics[width=0.8\textwidth]{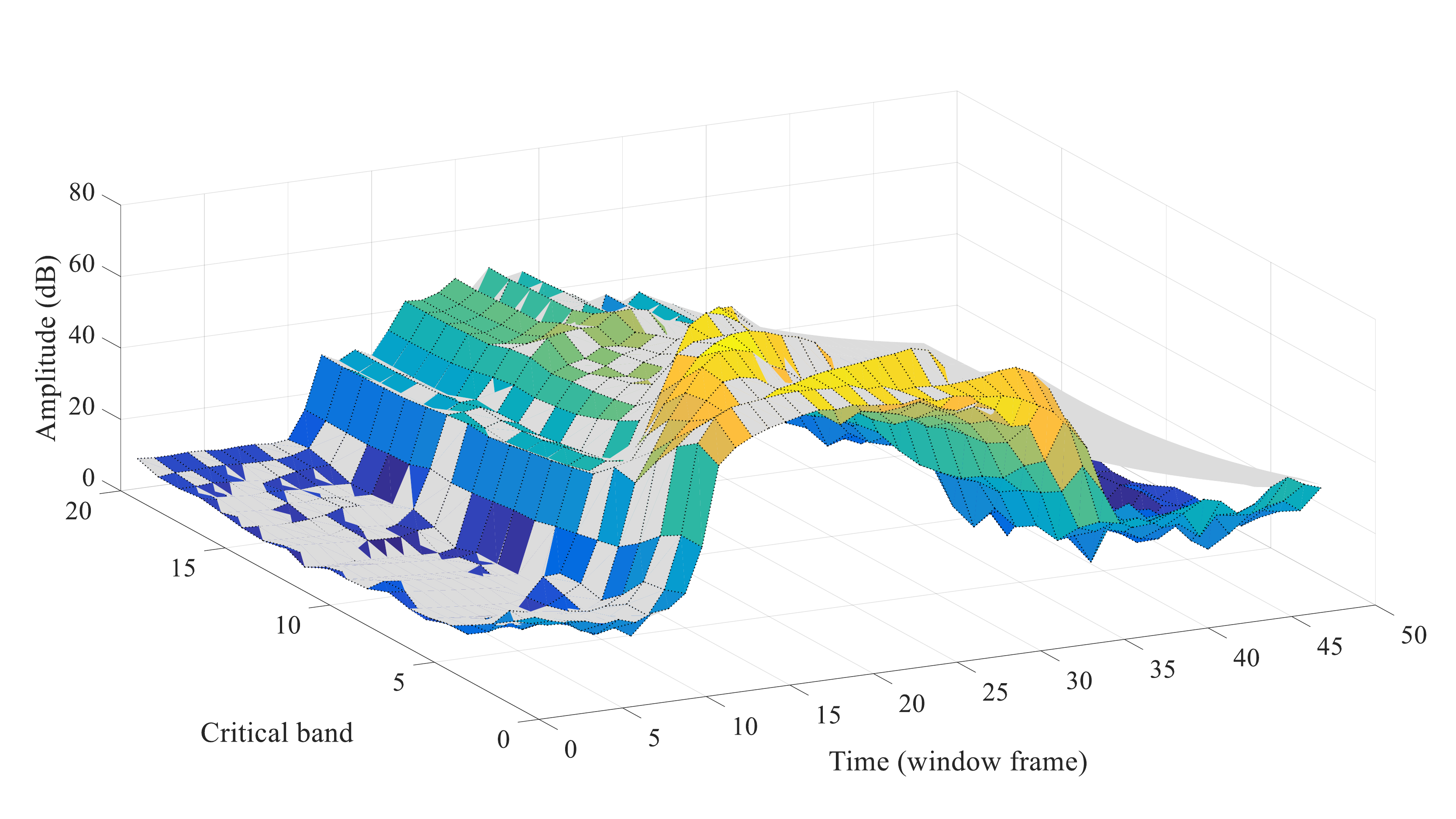}
    \caption{Both the simultaneous and temporal masking effects acting on the 3-D plot spectrogram of a speech utterance of ``one''. The grey-color shaded parts of the spectrogram are masked.}
    \label{All mask 3D}
\end{figure}

Figure \ref{All mask 3D} demonstrates the auditory masking effects acting in both the frequency and time domains, on a speech utterance of ``one'' in TIDIGITS dataset. The colored surface represents the original spectrogram while the grey areas represent the spectral energy values that are being masked. For TIDIGITS datasets, nearly half of the acoustic events (frames) are removed according to our auditory masking strategy, which corresponds to the $55\%$ removal of PCM pulses in speech coding \citep{ambikairajah2001wideband}.

\subsection{Cochlear filters and spike coding}
\label{spectral featuers and spike coding}
The human auditory system is primarily a frequency analyzer \citep{tobias2012foundations}. Many studies have confirmed the existence of the perceptual centre frequencies and equivalent bandwidths. To emulate the working  of the human cochlea, several artificial cochlear filter banks have been well studied: GammaTone filter bank \citep{patterson1987efficient}\citep{hohmann2002frequency}, Constant Q Transform-based filter bank (CQT) \citep{brown1991calculation}\citep{brown1992efficient}, Bark-scale filter bank \citep{smith1999bark}, etc. They share the same idea of logarithm distributed centre frequencies and constant Q factors but slightly differ in the exact parameters. To build the auditory encoding system, we adopt an event-based CQT-based filter bank in the time domain, following our previous work \citep{pan2018event}.

\subsubsection{Time-domain cochlear filter bank}

\begin{figure}[H]
    \centering
    \includegraphics[width=.8\columnwidth]{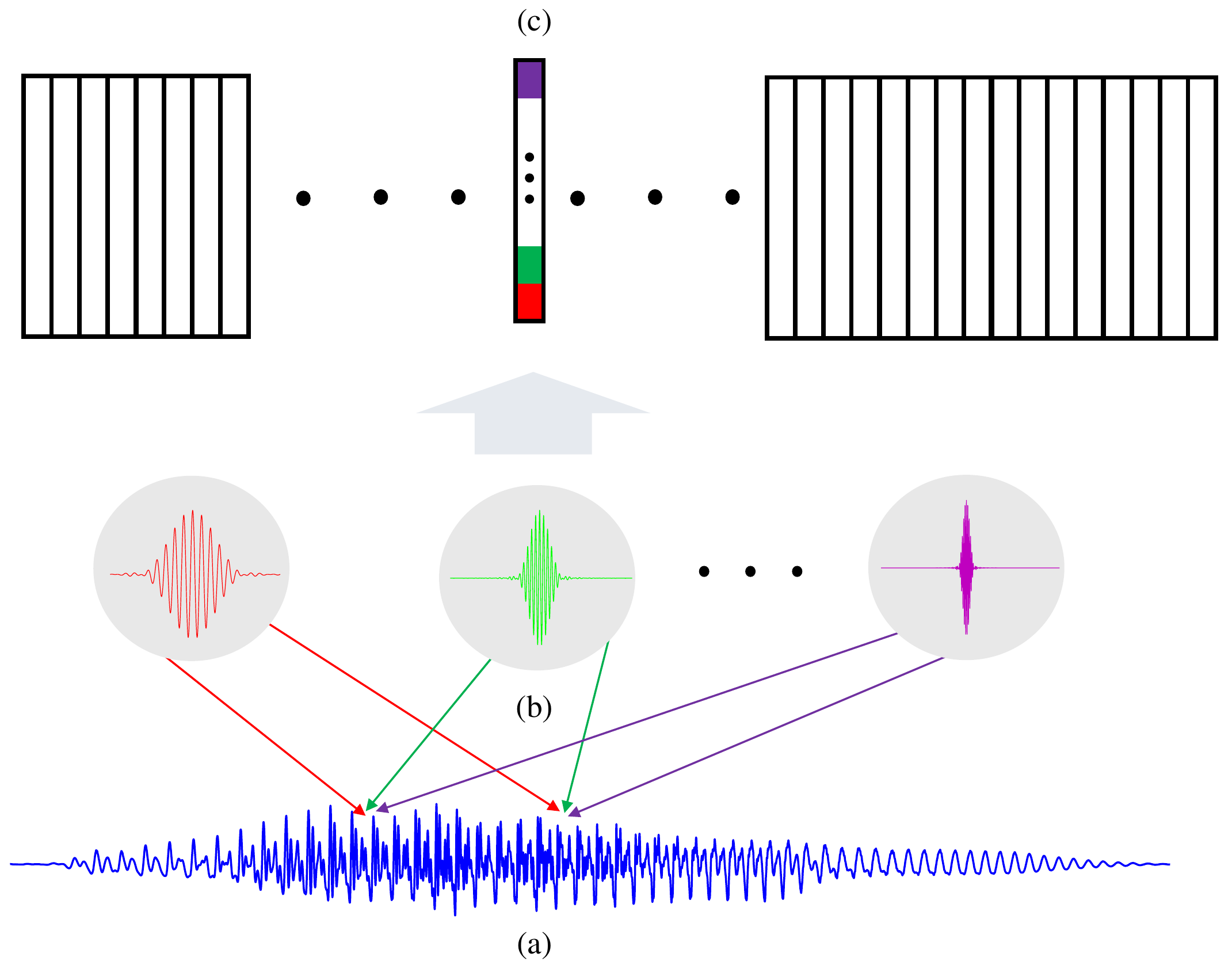}
    \centering
    \caption{(a) A speech signal of $M$ samples; (b) Time-domain filter bank with $K$ neurons that act as filters; (c) The output spectrogram that has $K \times M$ dimension}
    \label{Time convolution}
\end{figure}

Adopting an event-based approach to emulate the human auditory system, we propose a neuronal implementation of the event-driven cochlear filter bank, of which the computation can be parallelized as follows,

\begin{itemize}
\item As illustrated in Figure \ref{Time convolution}, a speech waveform (a) is filtered by $K$ neurons (b) where each neuron represents one cochlear filter from a particular frequency bin.

\item The weights of each neuron in (b) are set as the time-domain impulse response of the corresponding cochlear filter. The computing of a neuron with its input is inherently a time-domain convolution process. 

\item The output of the filter bank neurons is a $K$-length vector (c), where $K$ is the number of filters, for each time step. Since the signal (a) shifts sample by sample, the width of the output matrix is the same as the length of the input signal. As such, the auditory signal is decomposed into multiple channels in parallel, forming a spectrogram.
\end{itemize}

Suppose a speech signal $\mathbf{x}$ with $M$ samples $\mathbf{x}=[x_1,x_2,...,x_{M}]$ sampled at 16kHz. For the $k^{th}$ cochlear filter, the impulse response (wavelet) is a $M_k$-length vector $\mathbf{F}_k=[F_k(1),F_k(2),...,F_k(M_k)]$. We note the impulse response $\mathbf{F}_k$ has an infinite window size, however, numerically its amplitude decreases to small values outside an effective window, thus having little influence on the convolution results. As investigated in \citep{pan2018event}, we empirically set $M_k$ to an optimal value. So the $m^{th}$ output of the $k^{th}$ cochlear filter neuron is modeled as $y_{k}(m)$:
\begin{equation}
y_{k}(m)=\sum_{i=1}^{M_k} \phi_m(i)F_k(i),~k=1,2,...,K,~m=1,2,...,M
\end{equation}

\begin{equation}
{\phi}_m=[x_m,x_{m+1},x_{m+2},...,x_{m+M_k-1}],~ m \in 1,...,M
\end{equation}

${\phi}_m$ is a subset of the input samples within the $m^{th}$ window, whose length is the same as that of the $M_k$-length wavelet, indicated as the samples between the two arrows in Figure \ref{Time convolution} (a) and (b). The window ${\phi}_m$ will move sample by sample, naturally along with the flow of the input signal samples. At each time step, a vector of length $K$, which is the number of filters, is generated as shown in (c).  After $M$ such samples, the final output time-frequency map of the filter bank is a $K \times M$ matrix $\mathbf{Y}_{K \times M}$.

After time-domain cochlear filtering, the $K \times M$ time-frequency map $\mathbf{Y}_{K \times M}$ should be framed, which emulates the signal processing of hair cells in the auditory pathway. For the output waveform from each channel, we apply a framing window of length $l$ (samples) with a step size of $l/2$ and calculate the logarithmic frame energy $e$ of one framing window:

\begin{equation}
e=10\log(\sum_{q=1}^{l} x_q^2)
\label{eq:framing}
\end{equation}
where $x_q$ denotes the samples within the $l$-length window; $e$ is the spectral energy of one frame, hence obtaining the time-frequency spectrum $\mathbf{S}_{K \times N}(s_{ij})$ as indicated in Section \ref{sec:masking in both domains} which will be further encoded into spikes.

\subsubsection{Neural spike encoding}

In the inner ear, the motion of the stereocilia in the inner hair cells is converted into a chemical signal that excites adjacent nerve fibers, generating neural impulses that are then transmitted along the auditory pathway. Similarly, we would like to convert the sub-band framing energy into electrical impulses, or so-called spikes, for the purpose of information encoding and transmission. In prior work, the temporal dynamic sequences are encoded using several different methods: latency coding \citep{wu2018biologically}, phase coding \citep{arnal2012cortical}\citep{giraud2012cortical}, latency population coding \citep{dean2005neural}, that are adopted for specific applications. These encoding schemes are not optimized for SNN computation. 

We would like to propose a biologically plausible neural encoding scheme taking into account the three criteria as defined in Section \ref{sec: introduce}. In this section, the particular neural temporal coding scheme, which converts perceptual spectral power to precise spike times, is designed to meet the need of synaptic learning rules in SNNs \citep{gutig2006tempotron}\citep{ponulak2010supervised}. As such, the resulting temporal spike patterns are supposed to be friendly towards temporal learning rules.

In our previous work \citep{pan2019neural}, two mainstream neural coding schemes, the single neuron temporal codes and (latency coding, phase coding) and the population codes (population latency/phase coding, threshold coding) are compared. It is found that the threshold coding outperforms the other coding schemes in SNN-based pattern recognition tasks. Next are some observations made whilst comparing threshold coding, and the single neuron temporal coding.

First of all, the single temporal coding scheme, such as latency or phase coding, encodes the spectral power using spike delaying time, or phase-locking time. Suppose a frame of normalized spectral power is $e$, the $n^{th}$ latency spike timing $t^f_n=$ is defined as:

\begin{equation}
    t^f_n=(1-e)*T+(n-1)*T=(n-e)*T
    \label{latency code definition}
\end{equation}

where $T$ denotes the time duration of the encoding window. For the phase coding, $t^f_n$ is phase-locked to the nearest peak of the sub-threshold membrane oscillation. The spectral power, that represents the amplitude information, $e$ is represented as the relative spike timing $(1-e)*T$ within each window and the number of spikes embedded are in the order $n$. Unfortunately, the SNN can hardly decode such an encoding scheme without the knowledge of the encoding window boundaries, implicitly provided by the spike order $n$ and window length $T$. The spatio-temporal spike patterns could not provide such knowledge explicitly to the SNN. On the other hand, in the population code, such as threshold coding, the multiple encoding neurons naturally represent the amplitudes of the spectral power frames, and we only need to represent the temporal information in the spike timing. For example, the spike timing of the $n^{th}$ onset encoding neuron of the threshold code $t_f^n$ is:

\begin{equation}
	t_f^n=t_{\text{crossing}}
\end{equation}

$t_{\text{crossing}}$ records the time when the spectral tuning curve from one sub-band crosses the onset threshold $\theta_{n}$ of the $n^{th}$ encoding neuron. In this way, both the temporal and amplitude information is encoded and made known to the SNN, which meets the third criterion mentioned above.

Secondly, coding efficiency, which refers to the average encoding spike rates (number of spikes per second), is also studied in \citep{pan2019neural}. The threshold code has the least average spike rates among all investigated neural codes. As the threshold code encodes only threshold-crossing events, it is supposed to be the most efficient coding method.

Thirdly, the threshold code promises to be more robust against noise, such as spike jitter. As it encodes the trajectory of the dynamics of the sub-band spectral power, the perturbation of precise spike timing will have less impact on the sequence of encoding neurons.

As such, the threshold code is a promising encoding scheme for temporal sequence recognition tasks \citep{pan2019neural}. Further evaluation will be provided later in the experiments. While we note that each neural coding scheme has its own advantages, we focus on how the encoding scheme may help subsequent SNN learning algorithms in a cognitive task in this paper. As such, we adopt the threshold code for all experiments in this paper.

\subsection{Biologically plausible auditory encoding (BAE) with masking effects}
\label{sec: integrate encoder}

\begin{figure}[H]
    \centering
    \includegraphics[width=1.0\columnwidth]{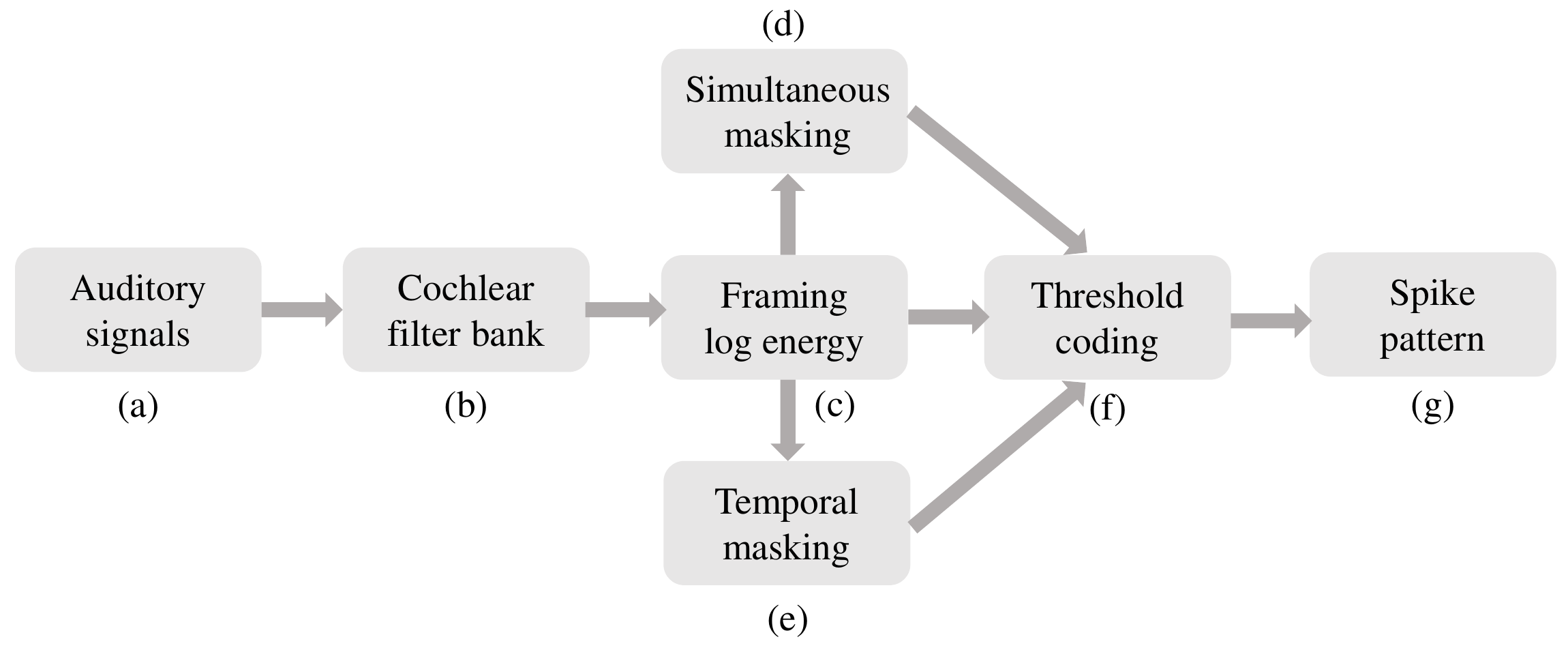}
    \centering
    \caption{The BAE scheme for temporal learning algorithms in auditory cognitive tasks. The raw auditory signals (a) are filtered by the CQT-based event-driven cochlear filter bank, resulting in a parallel stream of sub-band signals. For each sub-band, the signal is logarithmically framed, which corresponds to the processing in auditory hair cells. The framed spectral signals are then further masked in simultaneous and temporal masking. }
    \label{masking_block}
\end{figure}

We propose a BAE front-end with masking effects as illustrated in Figure  \ref{masking_block}.

Firstly the auditory stimuli are sensed and amplified by the microphone and some peripheral circuits, leading to a digital signal (a). This process corresponds to the pathway of the pinna, external auditory meatus, tympanic membrane and auditory tube. Then the physically sensed stimuli are filtered by the cochlear filter bank (b), that emulates the cochlear function of frequency analysis. The outputs of the cochlear filter bank are parallel streams of time-domain sub-band (or so-called critical band) signals with psycho-acoustic centre frequencies and bandwidths. For the purpose of further neural coding and cognitive tasks, the sub-band signals should be framed as the logarithm-scale energy as per equation \ref{eq:framing}. The output of (c), the raw spectrogram, is then converted into a precise spike pattern. The spectrogram is also being used to calculate the simultaneous and temporal masking levels, as in (d) and (e), under which the spikes will be omitted. Finally a sparse, perceptually related, and learnable temporal spike pattern for a learning SNN is generated as shown in (g).

\begin{figure}[H]
    \centering
    \includegraphics[width=1.0\columnwidth]{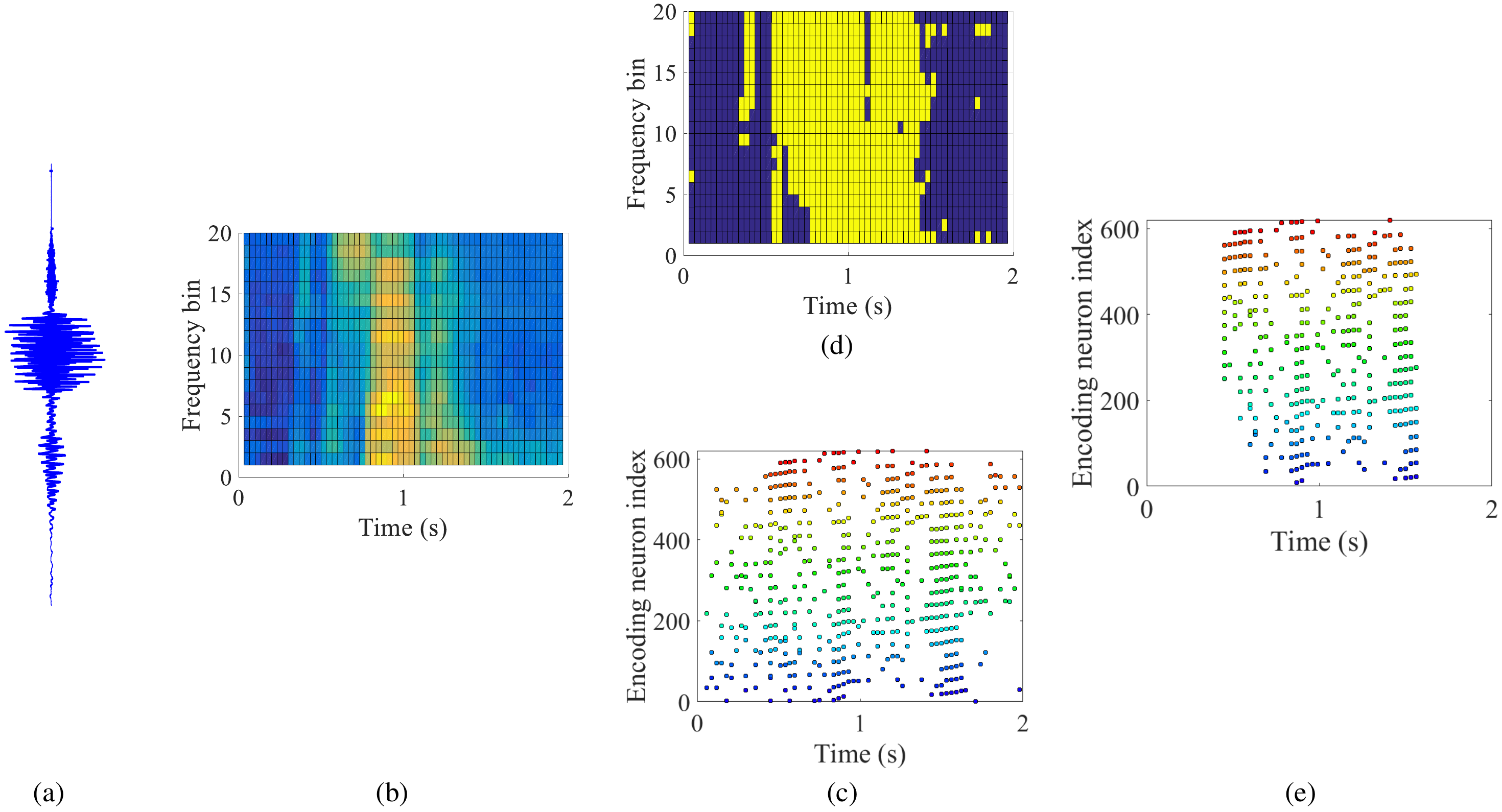}
    \centering
    \caption{An illustration of the intermediate results in a BAE process. Raw speech signal (a) of a speech utterance ``three'' is filtered and framed into a spectrogram (b), corresponding to the process in Figure \ref{masking_block}(b) and (c). By applying the neural threshold code, a precise spike pattern (c) is generated from the spectrogram. The masker map as described in Equation (4) is illustrated in (d), where yellow and dark blue color blocks represent value 1 and 0, respectively.  The masker (d) is applied on the spike pattern (c) and the auditory masked spike pattern is obtained in (e).}
    \label{Encode_block_demo}
\end{figure}

Figure \ref{Encode_block_demo} gives an example of the intermediate results at different stages in Figure \ref{masking_block} for a speech data waveform. Figure \ref{Encode_block_demo}(a) and (b) show the raw waveform and the spectrogram of a speech utterance ``three'' spoken by a male speaker. The spectrogram is further encoded into a raw spike pattern by threshold neural coding. Figure \ref{Encode_block_demo}(d) is the mask as formulated in Section \ref{auditory masking}, according to which the raw spike pattern \ref{Encode_block_demo}(c) is masked and results in a masked spike pattern (e). According to Table \ref{TIDIGITS speech quality measurements}, $50.48 \%$ of all spikes are discarded.

\section{Experiment and Results}
\label{experiment and results}

\subsection{Spike-TIDIGITS and Spike-TIMIT databases}
\label{sec:introduce datasets}

The TIDIGITS \citep{leonard1993tidigits} (LDC Catalog No. LDC93S10) is a speech corpus of spoken digits for speaker independent speech recognition  \citep{cooke2001robust} \citep{tamazin2019enhanced}. The speakers are from different genders (male and female), age ranges (adults and children), dialect districts (Boston, Richmond, Lubbock, etc.). As such, the corpus provides sufficiently speaker diversity and becomes one of the common benchmarking datasets. TIDIGITS has a vocabulary of 11 spoken words of digits. The original database contains both isolated digits and digit sequences. In this work, we only use the isolated digits: each utterance contains one individual spoken digit. In this first attempt, we would like to build a spike-version speech dataset that contains sufficient diversity and can be immediately used to train a SNN classifier \citep{wu2018biologically} \citep{pan2018event}. As each digit is repeated 224 and 226 times, the Spike-TIDIGITS has $224\times11=2464$ and $226\times11=2486$ isolated digit utterances for the  training and testing set, respectively.

The BAE encoder proposed in Section \ref{sec: integrate encoder} and Figure \ref{masking_block} is applied as the standard encoding scheme to generate this spike dataset. Table \ref{tab:TIDIGITS parameters} and Table \ref{tab:cochlear parameter} describe the parameters in the encoding process of Spike-TIDIGITS.

Next, we encode one of the most popular speech dataset TIMIT \citep{garofolo1993timit} into a spike-version, Spike-TIMIT. TIMIT dataset consists of richer acoustic-phonetic content than TIDIGITS  \citep{messaoud2011combining}. It consists of continuous speech utterances, that are useful for the evaluation of speech coding schemes \citep{besacier2000gsm}, speech enhancement \cite{el2007evaluation} or automatic speech recognition systems \citep{mohamed2011acoustic} \citep{graves2013speech}. Similar to TIDIGITS, the speakers of TIMIT corpus are from 8 different dialect regions in the United States, 438 males and 192 females. There are 4621 and 1679 speech sequences in the training and testing sets. This corpus has a vocabulary of 6224 words, which is larger than that of TIDIGITS. 

Our proposed BAE scheme is next evaluated in the following sections, using both reconstruction and speech pattern recognition experiments.

\subsection{Audio reconstruction from masked patterns}
\label{sec: recosntruction and PESQ}

According to equation \ref{eq: masking map}, we adopt the binary auditory mask $\Phi_{K \times N}({\phi}_{ij})$ which either fully encodes or ignores an acoustic event. It is suggested in auditory theory \citep{ambikairajah1997auditory} that partial masking may exist in the frequency domain, especially in the presence of rich frequency tones. We would like to evaluate the masking effect in the BAE front-end through both objectively and subjectively .

We begin by reconstructing the spike trains into speech signals, and then evaluate the speech quality using several objective speech quality measures: Perceptual Evaluation of Speech Quality (PESQ), Root Mean Square Error (RMSE) and Signal to Distortion Ratio (SDR). The PESQ, defined in \citep{beerends2002perceptual} \citep{rix2002perceptual}, is standardized as ITU-T recommendation P.862 for speech quality test methodology  \citep{recommendation2001perceptual}. The core principle of PESQ is the use of human auditory perception model \citep{rix2001perceptual} for speech quality assessment. For speech coding, especially the perceptual masking proposed in this paper, the PESQ measure could correctly distinguish between audible and inaudible distortions and thus assess the impact of perceptually masked coding noise. Besides, the PESQ is also used in the assessment of MPEG audio coding where auditory masking is involved.  In this paper, the PESQ scores are further converted to MOS-LQO (Mean Opinion Score-Listening Quality Objective) scores ranging 1 to 5, which are more intuitive for assessing speech quality. The mapping function is obtained from ITU-T  Recommendation P.862.1\citep{itu2003862}. Table \ref{tab:MOS score} defines the MOS scales and their corresponding speech quality subjective descriptions \citep{rec1996p}.

Besides PESQ, the RMSE (equation \ref{eq: RMSE}) and Expand SDR (equation \ref{eq: SDR}) measures are also reported, where ${x}_i$ and $\hat{x}_i$ denote the $i^{th}$ time-domain sample of the original and reconstructed speech signals $\boldsymbol{x}_{1 \times M}$ and $\hat{\boldsymbol{x}}_{1 \times M}$, respectively.

\begin{equation}
    \text{RMSE}=\sqrt{\frac{1}{M} \sum_{i=1}^M ({x}_i-\hat{x}_i)^2}
    \label{eq: RMSE}
\end{equation}

\begin{equation}
    \text{SDR}=10\log_{10} (\frac{\sum_{i=1}^M ({x}_i)^2}{\sum_{i=1}^M ({x}_i-\hat{x}_i)^2} )
    \label{eq: SDR}
\end{equation}

For comparison, we compare three groups of reconstructed speech signals: (1) the reconstructed signal $\hat{\boldsymbol{s}}_\text{mask}$ from spike trains with auditory masking; (2) the reconstructed signal $\hat{\boldsymbol{s}}_\text{raw}$ from raw spike trains without auditory masking; (3) the reconstructed signal $\hat{\boldsymbol{s}}_\text{random}$ from randomly masked spike trains.  

\begin{figure}[H]
    \centering
    \includegraphics[width=1.0\columnwidth]{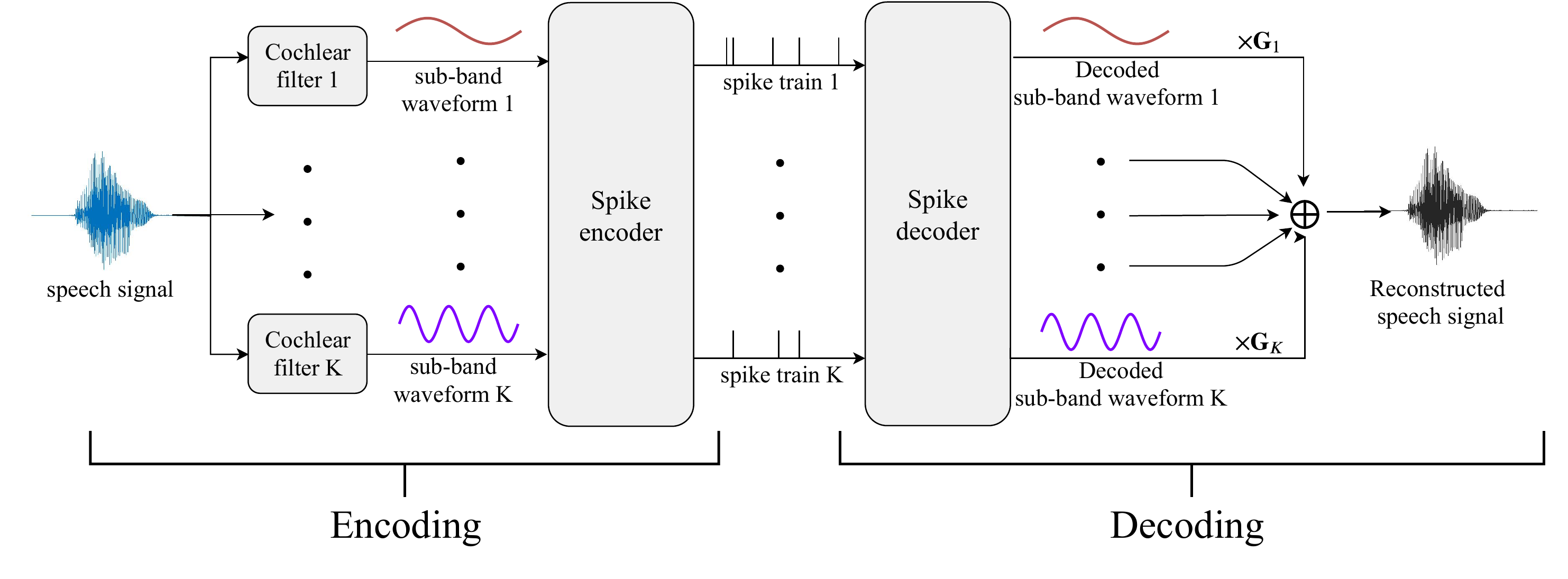}
    \centering
    \caption{The reconstruction from a spike pattern into a speech signal. Parallel streams of threshold-encoded spike trains that represent the dynamics of multiple frequency channels are first decoded into sub-band digital signals. The sub-band signals are further fed into a series of synthesis filters, which are built inversely from the corresponding analysis cochlear filters as in Figure \ref{Time convolution}. The synthesis filters compensate the gains from the analysis filters for each frequency bin. Finally, the outputs from the synthesis filter banks sum up to generate the reconstructed speech signal.
    }
    \label{synthesis_block}
\end{figure}

Figure \ref{synthesis_block} depicts the flowchart of the reconstruction process. The left and right panels represent the spike encoding and decoding processes. The raw speech signals are first decomposed by a series of cochlear analysis filters, generating parallel streams of sub-band signals as in Figure \ref{masking_block}(b). The 20 sub-band waveforms are encoded into spike times with masking strategies and then decoded back to sub-band speech signals. The reproduced sub-band waveforms 1 to $K$ (20 in this work) are gain-weighted and summed to form the reconstructed speech signal for perceptual quality evaluation. Since the cochlear filters decompose the input signal by various weighting gains in different frequency bands, the weighting gains in the decoding part represent the inverse processing of the cochlear filters.

The speech quality of the three groups of reconstructed signals is measured, as reported in Table \ref{TIDIGITS speech quality measurements} and \ref{TIMIT speech quality measurements}. For a fair comparison, we also simulate a random masking effect by dropping the same amount of spikes as that of the auditory masking. The raw spike patterns without any masking are used as a reference. The perceptual quality scores of the $\hat{\boldsymbol{s}}_\text{mask}$ and $\hat{\boldsymbol{s}}_\text{raw}$ are rather close at a high level of around 4.5, which suggests satisfying subjective quality between ``Excellent'' and ``Good'' according to Table \ref{tab:MOS score}. It is noted that the speech signals with random masking are perceived as ``Poor'' in quality. Besides the PESQ, the other two measures also lead to the same conclusion. The RMSE of $\hat{\boldsymbol{s}}_\text{raw}$ and $\hat{\boldsymbol{s}}_\text{mask}$ are approximately two orders of magnitude larger than that of the $\hat{\boldsymbol{s}}_\text{random}$; the SDRs also show a great gap.

\subsection{Speech recognition by SNN for TIDIGITS dataset}
\label{tempotron TIDIGITS}

In this section, we evaluate the BAE scheme in an SNN-based pattern recognition task, which also aims to evaluate the coding fidelity of our proposed methodology. The spike patterns encoded from TIDIGITS speech dataset are fed into an SNN, and the outputs correspond to the labels of which spoken digits the patterns are encoded from. The synapse efficacy updating rule is the MPD-AL, which is an efficient membrane potential driven aggregate-label learning algorithm for leaky integrate-and-fire spiking neurons \citep{MPDAL}. The network structure is given in Table \ref{tab: Tempotron structure}.

To evaluate the effectiveness of the BAE front-end, we compare the classification performances between spike patterns with and without auditory masking. Gaussian noise, measured by Signal-to-Noise Ratio (SNR) in dB, is added to the original speech waveforms before the encoding process. Table \ref{gaussian noise TIDIGITS} shows the classification accuracies under noisy conditions and in the clean condition.

The results show that the pattern classification accuracies of masked patterns are slightly higher than those of the original patterns, under different test conditions. Above all, referring to Table \ref{TIDIGITS speech quality measurements}, our proposed BAE scheme helps to reduce nearly half of the spikes, which is a dramatic improvement in coding efficiency.

\subsection{Large vocabulary speech recognition for TIMIT dataset}
\label{TIMIT results}
In Section \ref{sec: integrate encoder}, we present how the TIMIT dataset has been encoded into spike trains, which we henceforth refer to as Spike-TIMIT. We next train a recurrent neural network, the LSTM \citep{hochreiter1997long} on both the original TIMIT and Spike-TIMIT datasets, with the CTC loss function \citep{graves2006connectionist}. For the validation datasets, the normalized Levenshtein distance (by the labels) or the label error rate (LER) is reported \citep{graves2006connectionist}. We obtained an LER of 0.27 and 0.28 respectively for the TIMIT and Spike-TIMIT datasets. The network architecture of the LSTM used for both datasets is illustrated in Table \ref{tab: LSTM structure}. The LSTM networks are adapted from Tensorpack \citep{zhou2016dorefa}. We notice some improvement in accuracy when dropout is introduced for Spike-TIMIT but not for TIMIT. We further note that the Spike-TIMIT system involves many more input neurons than the TIMIT system (620 vs 39). However, because the TIMIT system employs more LSTM neurons, the Spike-TIMIT systems have much fewer parameters than the TIMIT system (4.5M vs 13M). This is also a desired outcome of the BAE front-end, that is, more neurons are used for neural encoding so that far less neurons and parameters are needed in the feature representation and classification pipeline, leading to overall saving in number of neurons and parameters.

\section{Discussion}
\label{discussion}

In this paper, we propose a biologically plausible auditory encoding (BAE) scheme,  especially for speech signals. The encoding scheme is inspired by the modeling of human auditory sensory system, which is composed of spectral analysis, neural spike coding, as well as the psycho-acoustic perception model. We adopt three criteria for formulating the auditory encoding scheme. 

For the spectral analysis part, a time-domain event-based cochlear filter bank is applied, with the perceptual scale of centre frequencies and bandwidths. The key feature of the spectral analysis is the parallel implementation of time-domain convolution. One of the most important properties of SNN is its asynchronous processing. The parallel implementation makes the neural encoding scheme a friendly front-end for any SNN processing. The neural encoding scheme,  the threshold code in our case, helps to generate a sparse and representative spike patterns for efficient computing in the SNN classifier. The threshold code helps in two aspects: firstly it tracks the trajectory of the spectral power tuning curves, which represents the features in the acoustic dynamics; secondly, the threshold code, as a form of population neural code, is able to project the dynamics in the time domain onto the spatial domain, which facilitates the parallel processing of spiking neurons on cognitive tasks \citep{pan2019neural}. Another key component of the BAE front-end is the implementation of auditory masking that benefits from findings in human psycho-acoustic experiments. The integrated auditory encoding scheme fulfills the three proposed design criteria. We have evaluated our BAE scheme through signal reconstruction and speech recognition experiments giving very promising results. To share our study with the research community, the spike-version of TIDIGITS and TIMIT speech corpus, namely, Spike-TIDIGITS and Spike-TIMIT, will be be made available as benchmarking datasets.

\begin{figure}[H]
    \centering
    \includegraphics[width=.8\columnwidth]{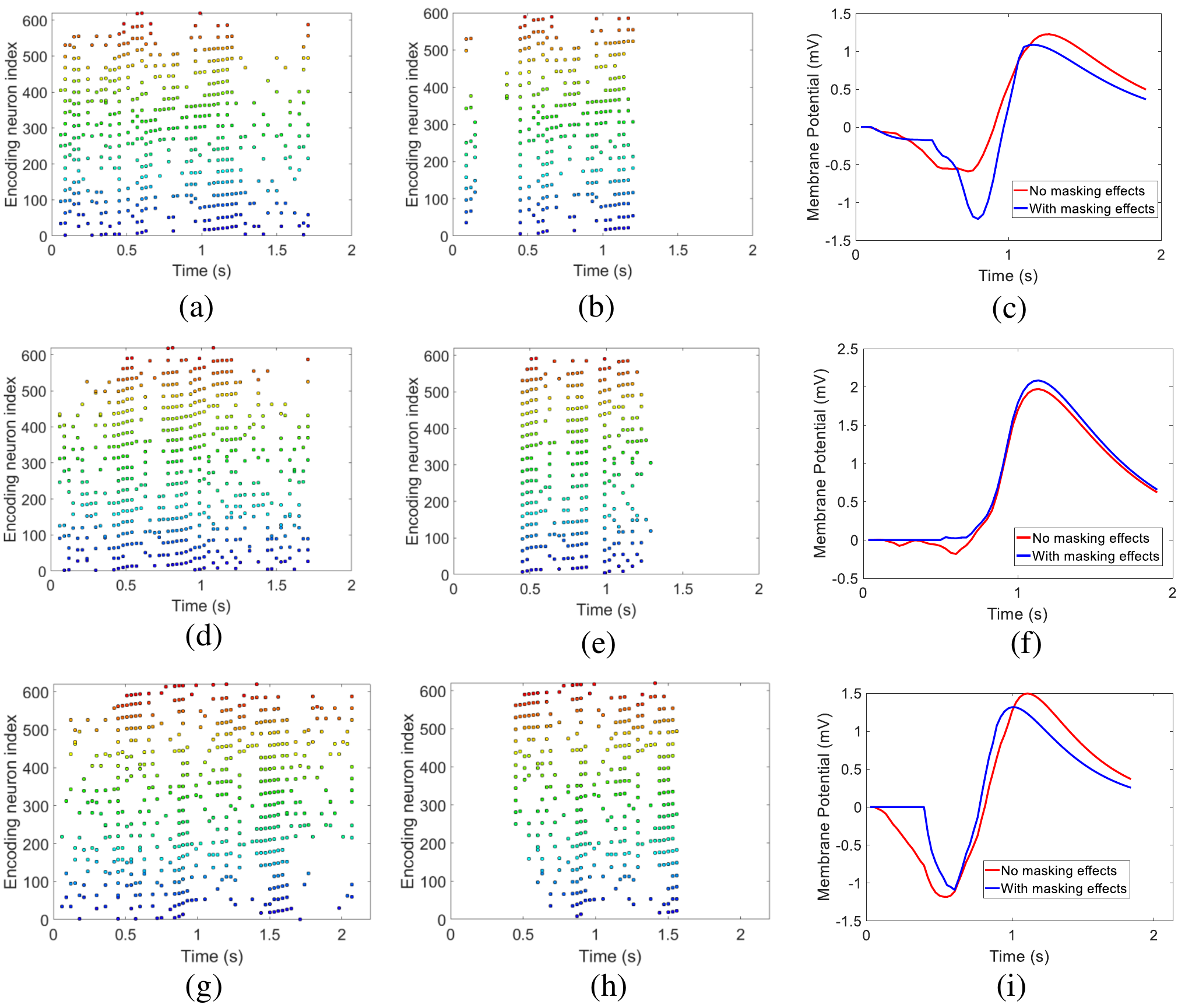}
    \centering
    \caption{Free membrane potential of trained Leaky-Integrate and Fire neurons, by feeding patterns with and without masking. The upper, middle, and lower panels are for three different speech utterances ``six'', ``seven'', and ``eight''. The spike patterns with or without masking are apparently different, but the output neuron follows similar membrane potential trajectories. }
    \label{membrane potential}
\end{figure}

Figure \ref{membrane potential} illustrates some interesting findings in our proposed auditory masking strategy. The upper, middle and lower panels of Figure \ref{membrane potential} represent three speech utterances from the TIDIGITS dataset. The first and second column illustrates the encoded spike patterns with and without auditory masking effects. It is apparent that a large number of spikes are removed. The graphs in the third column demonstrate the membrane potential of the output neuron in the trained SNN classifier after being fed with both patterns during the testing phase. For example, the LIF neuron in (c) responds to the speech utterance of ``six''. As such, the encoded pattern of spoken ``six'', as in (a) and (b) will trigger the corresponding neuron to fire a spike in the testing phase. The sub-figure (c) demonstrates that though the sub-threshold membrane potentials of masking/unmasking patterns have different trajectories, the two membrane potential curves will exceed the firing threshold (which is 1 in this example) at close timing. Similar results are observed in (f) and (i). The spike patterns with or without auditory masking lead to similar neuronal responses, either in spiking activities (firing or not) or in membrane potential dynamics, as observed in (c), (f), (i). It is interesting to observe that auditory masking has little impact on the neuronal dynamics. As a psycho-acoustic experiment, the auditory mask is always studied using listening tests. It remains unclear how the human auditory system responds to auditory masking. Figure \ref{membrane potential} provides an answer to the same question from a SNN perspective.

The parameters of auditory masking effects in this work, such as the exponential decaying parameter $c$ in equation \ref{eq: temporal mask}, or the cross-channel simultaneous masking thresholds in Figure \ref{Fre maskers}, are all derived in the acoustic model of MPEG-1 Layer III standard \citep{fogg2007mpeg} and tuned according to the particular tasks. However, from a neuroscience point of view, our brain is adaptive to different environments. This suggests that the parameters could be optimized by machine learning methodology, for different tasks and datasets. Also, the threshold neural code, which encodes the dynamics of the spectrum using threshold-crossing events, relies heavily on the choice of thresholds. We use 15 uniformly distributed thresholds for simplicity. We note that the recording of threshold-crossing events is analogous to quantization in digital coding, that the maximal coding efficiency (maximal information being conveyed constrained by the numbers of neurons or spikes) maybe derived using an information-theoretic approach. The Efficient Coding Hypothesis (ECH) \citep{barlow1961possible} \citep{srinivasan1982predictive} that describes the link between neural encoding and information theory could provide us the theoretical framework to determine the optimal threshold distribution in the neural threshold code. It may also otherwise be learned using machine learning techniques.

\section{Conclusion}
\label{conclusion}
Our proposed BAE scheme, motivated by the human auditory sensory system, could encode temporal speech data into spike patterns that are sparse, efficient, and friendly to SNN learning rules. It is both efficient and effective. We use the BAE scheme to encode popular speech datasets, namely, TIDIGITS and TIMIT into their spike versions: Spike-TIDIGITS and Spike-TIMIT. The two spike datasets are to be published as benchmarking datasets, in the hope of improving SNN-based classifiers.

\section*{Conflict of interest statement}
The authors declare that the research was conducted in the absence of any commercial or financial relationships that could be construed as a potential conflict of interest

\section*{Author contribution}
Zihan Pan performed the experiments and wrote the paper. All authors contributed to the experiments design, result interpretation and writing.

\section*{Funding}
This work was supported by in part by the Programmatic Grant No. A1687b0033 from the Singapore Government’s Research, Innovation and Enterprise 2020 plan (Advanced Manufacturing and Engineering domain).


\bibliographystyle{frontiersinSCNS_ENG_HUMS} 
\bibliography{test}

\section*{Figures}

\begin{figure}[H]
    \centering
    \includegraphics[width=1\columnwidth]{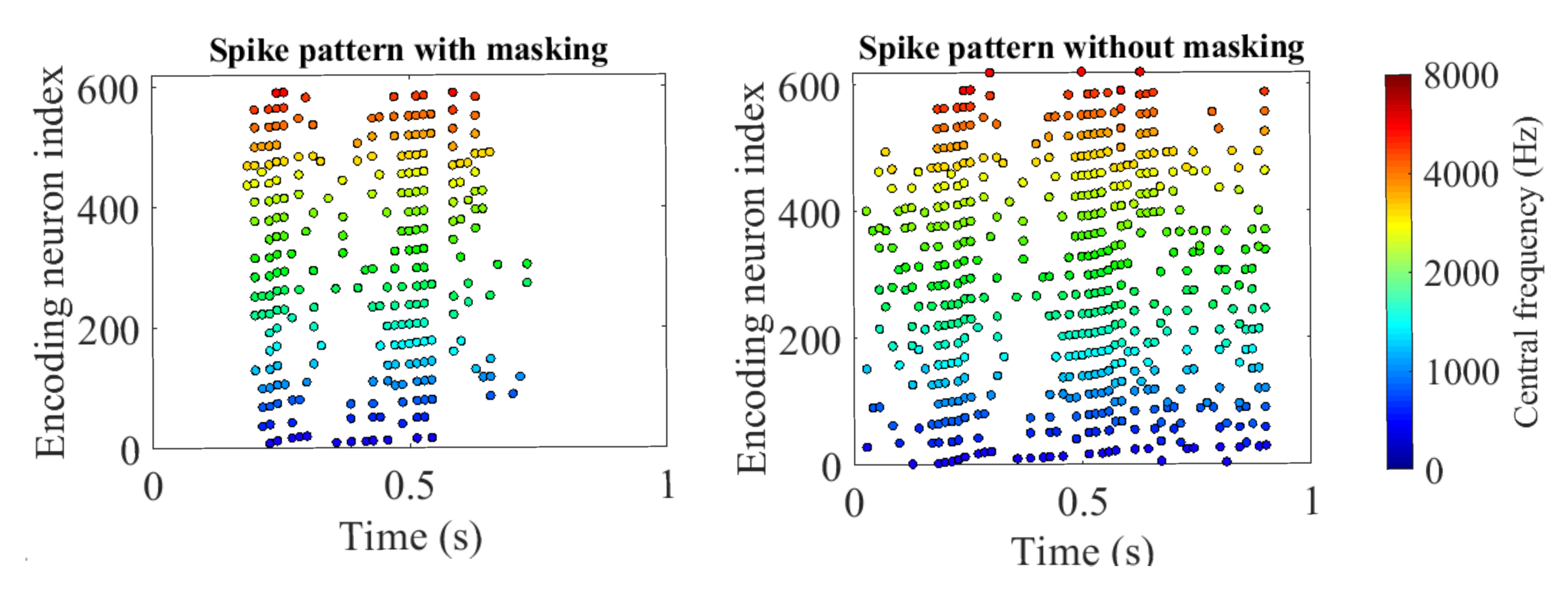}
    \centering
    \caption{Encoded spike patterns by threshold coding with/without masking. The two spike patterns are encoded from a speech utterance of ``five'' in TIDIGITS dataset. The x-axis and y-axis represent the time and encoding neuron index. The position of colorful dots indicate the spike timing of the corresponding encoding neurons. The colors distinguish the centre frequencies of the cochlear filter bank. With auditory masking, the number of spikes reduces by nearly $50\%$, which are close to the $55\%$ reducing rate of coding pulses as reported in \citep{ambikairajah1997auditory}.  }
    \label{compare pattern}
\end{figure}

\section*{Tables}

\begin{table}[H]
    \centering
    \begin{tabular}{c|c}
    \hline
     & parameters \\
    \hline 
         window size & $30$ms \\
         stride size &  $15$ms \\
         frequency range & $[200\text{Hz},8000\text{Hz}]$\\
         sampling rate & $20$kHz \\
    
    \end{tabular}
    \caption{Parameters of neural threshold encoding for TIDIGITS.}
    \label{tab:TIDIGITS parameters}
\end{table}

\begin{table}[H]
    \centering
    \begin{tabular}{c|cc}
    \hline
    Cochlear filter index & centre frequency (Hz) & bandwidth (Hz)\\ \hline
    1     &   200.2  & 69.3\\
    2     & 238.3 & 83.0\\
    3     & 283.2 & 98.6\\
    4     & 336.4& 117.2\\
    5     & 400.4& 139.6\\
    6     & 476.1& 166.0\\
    7     & 565.9& 197.3\\
    8     & 672.3& 234.4\\
    9     & 800.8& 278.3\\
    10     & 952.1& 331.1\\
    11     & 1131.3& 394.5\\
    12     & 1345.2& 468.8\\
    13     & 1600.6& 557.6\\
    14     & 1903.3& 663.1\\
    15     & 2263.7& 788.1\\
    16     & 2690.9& 937.5\\
    17     & 3200.2& 1114.3\\
    18     & 3805.7& 1325.2\\
    19     & 4525.9& 1576.2\\
    20     & 8000.5& 6949.2\\
    \end{tabular}
    \caption{Cochlear filter parameters: we use a total of 20 cochlear filters in the BAE front-end. The centre frequency and bandwidth of each filter are listed.}
    \label{tab:cochlear parameter}
\end{table}

\begin{table}[H]
    \centering
    \begin{tabular}{c|c|c|c|c|c}
    MSO scores &  5 & 4& 3& 2& 1\\ \hline
    Speech quality &  Excellent  & Good & Fair & Poor & Bad
    \end{tabular}
    \caption{MOS scales and their corresponding speech quality subjective assessments}
    \label{tab:MOS score}
\end{table}

\begin{table}[H]
    \centering
    \begin{tabular}{c|cccc}
    Reconstucted signals  &  PESQ & RMSE & SDR (dB) & Reduced rates ($\%$) \\ \hline
    
    $\hat{\boldsymbol{s}}_\text{raw}$ & $4.54$ & $4.78\times10^{-4}$ & $34.60$ & $0$ \\    
    
    $\hat{\boldsymbol{s}}_\text{mask}$ & $4.43$ & $7.49\times10^{-4}$ & $29.94$ & $50.48$ \\
    
    $\hat{\boldsymbol{s}}_\text{random}$ & $2.92$ & $1.05\times10^{-2}$ & $4.76$ & $49.91$ \\    
    
    \end{tabular}
    \caption{The objective speech quality meansures of the reconstructed speech signals for spoken digits TIDIGITS dataset. The reduced rates refer to the ratio of masked spikes.}
    \label{TIDIGITS speech quality measurements}
\end{table}

\begin{table}[H]
    \centering
    \begin{tabular}{c|cccc}
    Reconstructed signals  &  PESQ & RMSE & SDR (dB) & Reduced rates ($\%$) \\ \hline
    
    $\hat{\boldsymbol{s}}_\text{raw}$ & $4.54$ & $1.23\times10^{-4}$ & $42.28$ & $0$ \\    
    
    $\hat{\boldsymbol{s}}_\text{mask}$ & $4.44$ & $3.10\times10^{-4}$ & $34.02$ & $29.33$ \\
    
    $\hat{\boldsymbol{s}}_\text{random}$ & $2.35$ & $9.20\times10^{-3}$ & $4.83$ & $30.8$ \\    
    
    \end{tabular}
    \caption{The objective speech quality measures of the reconstructed speech signals for continuous and large vocabulary speech dataset TIMIT. The reduced rates refer to the ratio of masked spikes.}
    \label{TIMIT speech quality measurements}
\end{table}

\begin{table}[H]
\centering
\begin{tabular}{c|c|c|c}
\hline
Dataset & Input layer  & Output layer  \\ \hline
Spike-TIDIGITS & $1 \times 620$ encoding neurons & $1 \times 11$ Leaky Integrate-and-Fire neurons\\  
\end{tabular}%
\caption{SNN architectures for Spike-TIDIGITS classification}
\label{tab: Tempotron structure}
\end{table}

\begin{table}[H]
\centering
\begin{tabular}{c|cccccc}
\hline
SNR & -10 & 0 & 10 & 20 & 30 & clean\\ \hline
With masking & 59.5 & 78.2 & 87.5 & 91.9 & 93.5 & 97.4\\
W/o masking & \multicolumn{1}{c}{} 61.2 & 76.5 & 87.1 & 90.8 & 93.4 & 96.9
\end{tabular}%
\caption{TIDIGITS classification accuracies under Gaussian noise }
\label{gaussian noise TIDIGITS}
\end{table}

\begin{table}[H]
\centering
\begin{tabular}{c|c|c|c}
\hline
Dataset & Input layer & Hidden layer & Output layer  \\ \hline
TIMIT & $1 \times 39$ & $1\times 1024$ LSTM-$1\times 1024$ LSTM & $1\times 620$  \\
Spike-TIMIT & $1 \times 620$ & Dropout ($0.2$)-$1\times 512$ LSTM-$1\times 512$ LSTM & $1\times 620$  
\end{tabular}%
\caption{LSTM architectures for TIMIT and Spike-TIMIT classification}
\label{tab: LSTM structure}
\end{table}

\end{document}